\documentclass[12pt, reqno]{article}

\usepackage{jheppub}
\usepackage{amssymb}
\usepackage{amsmath}
\usepackage[usenames,dvipsnames]{xcolor}
\usepackage{epsfig}
\usepackage{dcolumn}
\usepackage{tikz}
\usetikzlibrary{shapes.geometric, arrows}
\usepackage{upgreek}
\usepackage{setspace}
\usepackage{enumitem}
\usepackage{array,multirow,bigdelim,arydshln}
\usepackage{appendix}
\usepackage{xparse}
\usepackage{mathtools}
\usepackage{hyperref}
\hypersetup{
	colorlinks,
	urlcolor=Maroon,
	linkcolor=Maroon,
	citecolor=Maroon
}
\NewDocumentCommand{\binomial}{omm}
 {%
  \genfrac(){0pt}{}{#2}{#3}%
  \IfValueT{#1}{_{\!#1}}%
 }
\NewDocumentCommand{\eulerian}{omm}
 {%
  \genfrac<>{0pt}{}{#2}{#3}%
  \IfValueT{#1}{_{\!#1}}%
 }

\newcommand{\overbar}[1]{\mkern 1.5mu\overline{\mkern-1.5mu#1\mkern-1.5mu}\mkern 1.5mu}
\usepackage{latexsym}
\usepackage{tikz}
\title{Extensions of Theories from Soft Limits}
\author[a]{Freddy Cachazo,}
\author[a,b]{Peter Cha}
\author[a,b]{and Sebastian Mizera}
\affiliation[a]{Perimeter Institute for Theoretical Physics, Waterloo, ON N2L 2Y5, Canada}
\affiliation[b]{Department of Physics $\&$ Astronomy, University of Waterloo, Waterloo, ON N2L 3G1, Canada}
\emailAdd{fcachazo, pcha, smizera@pitp.ca}
\arxivnumber{1604.03893}

\abstract{We study a variety of field theories with vanishing single soft limits. In all cases, the structure of the soft limit is controlled by a larger theory, which provides an extension of the original one by adding more fields and interactions. Our main example is the $U(N)$ non-linear sigma model in its CHY representation. Its extension is a theory in which the NLSM Goldstone bosons interact with a cubic biadjoint scalar. Other theories we study and extend are the special Galileon and Born-Infeld theory, including its maximally supersymmetric version in four dimensions, the DBI-Volkov-Akulov theory. In all the cases, we propose the CHY representation of the complete tree-level S-matrix of the extended theories. In fact, CHY formulas are the key technique for studying the single soft limit behavior of the original theories. As a byproduct, we show that the tree-level S-matrix of the extended NLSM theory can be constructed using a very compact BCFW-like recursion relation, where physical poles are at most linear in the deformation parameter.
}

\begin{document}
\maketitle
\def \tr {\nonumber\\}
\def \la  {\langle}
\def \ra {\rangle}
\def\hset{\texttt{h}}
\def\gset{\texttt{g}}
\def\sset{\texttt{s}}
\def \be {\begin{equation}}
\def \en {\end{equation}}
\def \bes {\begin{eqnarray}}
\def \ens {\end{eqnarray}}
\def \k {\kappa}
\def \h {\hbar}
\def \r {\rho}
\def \l {\lambda}
\def \pf {\mathrm{Pf}}
\def \red {\color{red}}
\DeclarePairedDelimiter\floor{\lfloor}{\rfloor}
\def \A {\textsf{A}}


\numberwithin{equation}{section}

\section{Introduction}

The vanishing of scattering amplitudes in effective field theories of Goldstone bosons when a single particle becomes soft is a very well-understood phenomenon. This is known as the ``Adler zero" \cite{Adler:1964um, Susskind:1970gf} and was discovered in the 60's during the study of pion interactions. The order of the vanishing of the amplitude $\tau$ has recently been used as a classification tool of effective field theories \cite{Cheung:2014dqa,Cheung:2015ota}. In this classification, several theories stand out as being the ones with the currently maximal soft behavior. These theories include the non-linear sigma model (NLSM), Born-Infeld (BI) and a special Galileon (sGal).

The standard lore is that when the single soft limit vanishes, the simultaneous double soft limit, i.e., the limit when two particles become soft at the same time, carries the relevant physical information about the spontaneously broken symmetries. The textbook example is the $U(N)$ NLSM. While single soft limits vanish in this theory, double soft limits are finite and depend on the relative directions of the momenta taken to be soft. This residual dependence has been the subject of a renewed wave of interest since 2008, when it was shown to contain the $E_{7(7)}$ non-linearly realized symmetry of ${\cal N}=8$ supergravity in four dimensions \cite{ArkaniHamed:2008gz}.

In this work, we show that the single soft limit of a variety of effective field theories also has a very interesting structure. In fact, it contains a whole new theory in it. The new theory can be thought of as an extension of the first one in which the original fields interact with new ones and where the flavor group is also enlarged.

Schematically, we write
\be\label{simple}
A_n^{\mathrm{theory}_1} \xrightarrow{\mathrm{soft\, limit}} \tau^p\, A_{n-1}^{\mathrm{theory}_1 \oplus\, \mathrm{theory}_2} + \mathcal{O}(\tau^{p+1}), \quad p > 0.
\en
Here theory$_1$ is the original theory and theory$_2$ refers to that of the new additional fields. The sum in $\mathrm{theory}_1 \oplus\, \mathrm{theory}_2$ indicates that interaction terms among fields from both theories are present. We call the larger theory the \emph{extension} of the original one.

In this work we initiate the study of these extensions. Our main example is the $U(N)$ NLSM, whose extension contains an additional $U(\tilde N)$ flavor group and a cubic biadjoint scalar theory. We also study the special Galileon whose extension contains a $U(N) \times U(\tilde N)$ biadjoint scalar, as well as one NLSM field for each of the flavor groups.

The structure of the single soft limit in \eqref{simple} is slightly oversimplified in that the right hand side is a sum over terms which ends up revealing the structure of a new flavor group. This surprisingly rich structure hiding in the soft limit can be easily discovered using the Cachazo-He-Yuan (CHY) representation  \cite{1307.2199 CHY,Cachazo:2013iea,Dolan:2013isa,CHY} of the corresponding original amplitudes.

The CHY formula expresses $n$-particle amplitudes as a contour integral over the space of $n$-punctured complex spheres. One of the main virtues of the CHY representation is that soft limits follow from a simple residue theorem argument and give rise to formulas which are themselves integrals over the space of $(n-1)$-punctured complex spheres. After some simple manipulations the new contour integrals turn out to have the properties of well-defined amplitudes. These are the ones identified with amplitudes of the extended theory.

It is important to notice that while the CHY formulation of the NLSM, DBI and sGal have passed many non-trivial tests \cite{CHY}, they remain proposals and hence the conclusions in this work are strictly speaking properties of these representations.

We also consider examples of theories with spin. These are theories describing the spontaneous symmetry breaking of space-time symmetries such as those arising from the effective theory of $D$-branes. The classic example is the Dirac-Born-Infeld theory. In this case, we find that the extension contains a $U(N)$ gauge group. In other words, the extended theory contains Yang-Mills gauge bosons interacting with the original abelian Born-Infeld fields.

The CHY formulations are tailored to space-times of arbitrary dimension. However, when fermions are included it is best to restrict the discussion to a fixed dimension. Four dimensions is the best understood case. In fact, CHY formulas reduce to Witten-RSV-like formulas \cite{Witten, RSV, Cachazo:2012kg, Cachazo:2012pz} and split into helicity sectors in theories with spin.

In the CHY formulation, the DBI theory is built out of two pieces. One of them carries all the information of polarization vectors. The other part is purely scalar in nature. Specializing to four dimensions allows us to supersymmetrize the former without modifying the latter. We propose that the maximally supersymmetric case corresponds to a DBI-Volkov-Akulov theory \cite{Volkov:1973ix} with sixteen linearly realized supercharges and sixteen non-linearly realized supersymmetries\footnote{After this work was completed and was being prepared for submission, the paper \cite{He:2016vfi} has appeared, where a similar proposal for constructing the S-matrix of the DBI-VA theory was introduced.}. The complete action for this theory was found in 2013 along with other formulas for less supersymmetric models in \cite{Bergshoeff:2013pia}.

An important byproduct of our analysis is a novel recursion relation for the $U(N)$ NLSM. BCFW-like recursion relations for this theory have been the subject of recent interest based on the idea of exploiting the soft structure to tame the behavior for large deformation parameters \cite{Cheung:2014dqa, Luo:2015tat}. Here, we exploit the fact that not only the single soft limit vanishes, but also that the coefficient of the zero is controlled by lower point amplitudes in the extended theory. We close the recursion by showing that general amplitudes in the extended theory can also be recursed from lower point ones. The complex deformation of momenta we employ is the minimal one, which allows for the use of the soft limit of a single particle, say the $n^{\rm th}$ particle. The deformation uses two other particles, $\{i,j\}$, and deforms $k_i(z)$, $k_j(z)$ in order to produce $k_n(z) = (1-z)k_n$. The deformation is done in such a way that all kinematic invariants are at most linear in $z$. When $i,j$ and $n$ are three consecutive labels of the flavor ordering, the recursion relation is very compact.

This paper is structured as follows. The main example of the non-linear sigma model extension is introduced in Section \ref{sec:NLSM}. There we also define linear on-shell recursion relations for this model. In Section \ref{sec:Galileon} we present the derivation of extensions of the special Galileon theory using KLT relations. Analogous derivation for Born-Infeld theory is shown in Section \ref{sec:BI}. Finally, by specializing to four dimensions we propose a formula calculating the tree-level S-matrix in DBI-Volkov-Akulov theory and find its extension in Section \ref{sec:DBI-VA}. We close with conclusions and outlook in Section \ref{sec:Discussion}.

\section{\label{sec:NLSM} Extension of the $U(N)$ Non-linear Sigma Model}

Non-linear sigma models (NLSM) originated as effective theories of pion scattering \cite{GellMann:1960np} and played a pivotal role in the development of chiral perturbation theory. It is a well-known fact that scattering amplitudes in these theories vanish in the single soft limit \cite{Adler:1964um,Susskind:1970gf}, indicating the fact that there is a non-linearly realized symmetry. Somewhat surprisingly, double soft limits are non-vanishing but are direction dependent. This structure contains physical information and has received attention relatively recently in the literature since the work \cite{ArkaniHamed:2008gz}.

In this section, we take a step back to carefully analyze the single soft limits of massless $U(N)$ NLSM in view of its CHY representation. In \cite{CHY} it was proposed that $n$-particle amplitudes in the theory can be computed as an integral over the moduli space of $n$-punctured spheres. This formulation makes it easy to see not only that the single soft limit vanishes, but also that it is as a linear combination of integrals over $(n-1)$-punctured spheres. In this section we find an interpretation for these new integrals as amplitudes in a theory that not only contains NLSM scalars, but also a whole new flavor group and a new sector of scalars transforming in the biadjoint representation with a cubic self-interaction.

We interpret this result as saying that the $U(N)$ NLSM is naturally part of a larger theory, which we call its extension. Having expressed the single soft limit in terms of lower point amplitudes in the extended theory, it is clear that a BCFW-like recursion relation could exist. In this section we also study this question and find a set of recursion relations which produce compact formulas for the amplitudes. Tree-level scattering amplitudes of $U(N)$ non-linear sigma model have been previously studied in \cite{Kampf:2013vha} and Berends-Giele like recursion relations have been found.

\subsection{Single Soft Limit}

In order to study the single soft limit let us review the CHY formulation of amplitudes in the $U(N)$ NLSM. Each NLSM scalar carries an adjoint index and amplitudes can be color decomposed
\be
{\cal A}_n^{\rm NLSM} = {\rm Tr} \left( T^{a_1}T^{a_2}\cdots T^{a_n}\right)A_n^{\rm NLSM}(1, 2, \ldots ,n) + \ldots
\en
The amplitude $A_n^{\rm NLSM}(1, 2,\ldots, n)$ is called a flavor-ordered partial amplitude. The total amplitude ${\cal A}_n^{\rm NLSM}$ is computed by summing over $(n-1)!$ terms corresponding to all possible orders modulo cyclic transformations. In order to simplify the notation we identify the canonical ordering with the identity permutation  $\mathbb{I}_n = (1, \dots, n)$.

The CHY representation of $A_n^{\rm NLSM}(\mathbb{I}_n)$ is given by \cite{CHY}
\be
A_n^{\rm NLSM: CHY}(\mathbb{I}_n) = \oint d\mu_n~\mathcal{C}_n (\mathbb{I}_n)~(\pf' \A_n)^2.
\en
The definition of the measure $d\mu_n$ is given in Appendix \ref{sec:Review CHY}, where a review of the CHY formulation is presented. All we need here is that it is a measure over the space of $n$ punctures on a sphere with locations denoted by $\sigma_1,\sigma_2,\ldots, \sigma_n$. The contour of integration computes the residue of the integral on the solutions of the equations
\be
E_a = \sum_{b=1,b\neq a}^n \frac{s_{ab}}{\sigma_{ab}} = 0, \quad {\rm with}\quad \sigma_{ab}\equiv \sigma_a-\sigma_b.
\en
These are known as the scattering equations \cite{Cachazo:2013gna}.

The two ingredients in the integrand are the so-called Parke-Taylor factor
\be
\mathcal{C}_n(\mathbb{I}_n) = \frac{1}{\sigma_{12}\sigma_{23}\cdots \sigma_{n1}}
\en
and the reduced Pfaffian of an $n\times n$ matrix $\A_n$ with components $s_{ab}/\sigma_{ab}$. The reduced Pfaffian is defined by $\pf' \A_n =\frac{(-)^{i+j}}{\sigma_{ij}}\pf\, \A_n^{[ij]}$. Here $\A_n^{[ij]}$ denotes the submatrix of $\A_n$ obtained by deleting the rows $i,j$ and columns $i,j$. The fact that $\pf' \A_n$ is independent of this choice follows from the scattering equations, which give $\A_n$ co-rank $2$.

The CHY formula $A_n^{\rm NLSM: CHY}(\mathbb{I}_n)$ has passed many non-trivial consistency checks and from now on we drop the superscript ``CHY" and simply treat it as the definition of the amplitudes of interest.

Now we are ready to study the leading order behavior of the CHY integrand in the soft limit, i.e., taking $k_n = \tau \hat{k}_n$ with $\tau \to 0$. Starting with
\be
\pf' \A_n \to \frac{\tau}{\sigma_{1,n-1}} \sum_{a=2}^{n-2}\, (-)^a\, \frac{\hat{s}_{an}}{\sigma_{an}}\; \pf\, \A_{n}^{[1,a,n-1,n]},
\en
where again the notation using square brackets means we have removed the columns and rows labeled by $1,a,n-1,n$ from the matrix $\A_n$. Next, we note that the Parke-Taylor factor satisfies  the following convenient identity:
\be
\mathcal{C}_n (\mathbb{I}_n) = \mathcal{C}_{n-1} (\mathbb{I}_{n-1})\; \frac{\sigma_{n-1,1}}{\sigma_{n-1,n}\, \sigma_{n,1}}.
\en

The $n^{\mathrm{th}}$ scattering equation can be trivially written as $E_n \to \tau \hat{E}_n$. All other scattering equations become the equations for a system on $n-1$ particles and independent of $\sigma_n$ when $\tau$ is taken to be zero.

We now have all the tools necessary to study the single soft limit. Since $\sigma_n$ only appears in $E_n$, one can single out the $\sigma_n$ contour integral
\bes
A^{\mathrm{NLSM}}_n (\mathbb{I}_n) &=& - \tau \oint d\mu_{n-1}\, \mathcal{C}_{n-1} (\mathbb{I}_{n-1})\tr
&&\qquad\times \oint_{\Gamma} \frac{d\sigma_n}{\hat{E}_n} \frac{1}{\sigma_{n-1,n}\, \sigma_{n,1}\, \sigma_{1,n-1}} \left( \sum_{a=2}^{n-2}\, (-)^a\, \frac{\hat{s}_{an}}{\sigma_{an}}\; \pf\, \A_{n}^{[1,a,n-1,n]} \right)^2 + \mathcal{O}(\tau^2),\nonumber
\ens
where the contour $\Gamma$ encloses the poles of $\hat{E}_n$ in the $\sigma_n$ complex plane. Since there is no pole at infinity, we can deform $\Gamma$ so that the residue is over all the remaining poles. There are poles only at $\sigma_n = \sigma_2, \sigma_3 ,\dots, \sigma_{n-2}$. We thus obtain the final result:
\be
\label{eq:NLSM extension}
\boxed{A^{\mathrm{NLSM}}_n (\mathbb{I}_n) = \tau\, \sum_{a=2}^{n-2} \hat{s}_{an}  A_{n-1}^{\mathrm{NLSM}\oplus\mathrm{\phi^3}}(\mathbb{I}_{n-1}\, |\, n-1,a,1) +\mathcal{O}(\tau^2),}
\en
where
\be
\label{eq:NLSM-phi}
A_n^{\mathrm{NLSM} \oplus \mathrm{\phi^3}}(\alpha | \beta) = \oint d\mu_n \; \Big( \mathcal{C}_n (\alpha)\Big) \; \Big( \mathcal{C}(\beta) \, (\pf\, \A_{\overbar{\beta}})^2 \Big).
\en
Our notation already hints that these new objects have the interpretation of a theory in which NLSM fields interacts with a cubic biadjoint scalar. As explained in the introduction, we use the symbol $\oplus$ to indicate the presence of interactions. In order to justify this interpretation, note that the left Parke-Taylor factor carries the $U(N)$ flavor information and it means that whatever new particles must be added, have to be in the adjoint of this group. The right half-integrand splits into a Parke-Taylor factor indicating the presence of biadjoint scalars and the Pfaffian containing the NLSM Goldstone bosons. Clearly, both species share the partial ordering $\alpha$, but the biadjoints obey the additional ordering $\beta$ in the $U(\tilde{N})$ indices. Here $\A_{\overbar{\beta}}$ is a shorthand for a $|\overbar{\beta}| \times |\overbar{\beta}|$ matrix $\A$ involving only the particles in the complement of the set $\beta$. Depending on the case we can use one of the two equivalent notations $\A_{\overbar{\beta}} = \A^{[\beta]}_n$.

We have thus discovered that despite NLSM amplitudes vanishing in the soft limit as $\mathcal{O}(\tau)$, there is an object we can associate to this limit. Even more, the object turns out to be a theory of non-linear sigma model scalars enriched with interactions with the biadjoint theory! We dub it the extension of the $U(N)$ NLSM.

Note that in the definition (\ref{eq:NLSM-phi}) it is understood that the Pfaffian is reduced when $\beta = \varnothing$, and equals zero whenever $|\beta|=1$. It is also evident that due to antisymmetry of the matrix $\A$, the amplitude with odd number of sigma fields vanishes. A curious property of this formulation is that an amplitude with $n$ NLSM particles is identical to an amplitude with $n-2$ sigmas and $2$ biadjoint scalars. We postpone listing of the examples of amplitudes in this extended theory until later in this section when we derive more tools for their explicit computation.

\subsection{Linear BCFW-like Recursion Relation}

The $U(N)$ NLSM, like many scalar theories, is not constructible via the standard BCFW construction \cite{Britto:2004ap,Britto:2005fq} due to the behavior of the amplitudes at large complex momenta. In order to overcome this problem, several procedures have been developed. One of them is using Berends-Giele recursion relations \cite{Kampf:2013vha}. Another one, more related to our construction, is to exploit the vanishing of soft limits in order to construct rational functions with better behavior for large momenta \cite{Cheung:2015ota}.

In the second class of techniques, Cheung, et al. introduced a modified method of defining BCFW-like recursion relations, where the momentum of each particle is deformed as $k_a\to (1 + z r_a)k_a$. This is in order to use the soft behavior of the NLSM, DBI, and Galileon theories in each leg \cite{Cheung:2015ota}. Under this momentum deformation, all kinematic invariants turn out to be quadratic polynomials of $z$, which means that each factorization channel receives contributions from two points in the complex $z$ plane.

In this section we show how the use of the extension of the $U(N)$ non-linear sigma model allows us to define a minimal recursion scheme with each internal propagator being linear in the deformation parameter.

Let us consider three external particles: $i,j,n$. In the four-dimensional space spanned by the corresponding momenta we can perform the following deformation\footnote{In this work we use the conventions $\la ab \ra = \epsilon_{\alpha \beta} \lambda^\alpha_a \lambda^\beta_b$, $[ab] = \epsilon_{\dot{\alpha} \dot{\beta}} \tilde{\lambda}^{\dot{\alpha}}_a \tilde{\lambda}^{\dot{\beta}}_b$ with Mandelstam invariants $s_{ab} = \la ab \ra [ab]$.}:
\be
\label{eq:3-shift}
\tilde{\lambda}_n \to \hat{\tilde{\lambda}}_n = (1-z) \tilde{\lambda}_n, \qquad \lambda_i \to \hat{\lambda}_i = \lambda_i + \alpha z \lambda_n, \qquad \lambda_j \to \hat{\lambda}_j = \lambda_j + \beta z \lambda_n.
\en
Imposing conservation of momentum at any value of $z$ gives the unique solution, $\alpha = \frac{[nj]}{[ij]}$, $\beta = \frac{[ni]}{[ji]}$. This deformation can be thought of as a composition of two BCFW deformations \cite{Britto:2004ap,Bern:2005hs}.
Of course, we could also have declared that $\lambda_n \to \hat{\lambda}_n = (1-z) \lambda_n$, since the theory under consideration only contains scalar particles. In this case, this is a special kind of a Risager deformation \cite{Risager:2005vk}. Both choices are completely equivalent and from now on we work with the one in \eqref{eq:3-shift}.

This defines the \emph{3-shift} $(ijn)$. The limit $z \to 1$ explores the soft limit of the amplitude. Moreover, it is manifest that all Mandelstam invariants---and hence also the propagators---are at most linear in $z$.

We can now apply the 3-shift to the non-linear sigma model. In Appendix \ref{sec:large z} we show that in the large $z$ limit the amplitude goes as $\mathcal{O}(z)$ whenever $i,j,n$ are taken to be consecutive. This behavior at large $z$ is usually taken as a sign that a recursion is not possible, since the pole at infinity does not have a simple expression in terms of lower point amplitudes. 

In our case we can circumvent this problem. We use the fact that a very precise knowledge of the soft limits is available in terms of the lower point amplitudes in order to construct a rational function that vanishes at infinity
\be
\frac{\hat{A}_n^{\rm NLSM}(\mathbb{I}_n;z)}{(1-z)^2}
\en
and whose poles and residues are all given in terms of smaller amplitudes. Of course, near the soft limit point $z\sim 1$ the lower point amplitudes are not the NLSM ones, but those of the extended theory!

Now consider the following contour integral representation of the amplitude,
\be
A^{\rm NLSM}_n(\mathbb{I}_n) = \oint_{|z|=\epsilon} dz\; \frac{\hat{A}^{\mathrm{NLSM}}_n(\mathbb{I}_n; z)}{z(1-z)^2}.
\en
The contour is a small circle around $z=0$. Using the residue theorem we can evaluate the integral by summing over all other poles of the function in the complex $z$ plane. There are only two types of poles: at $z=1$ we explore the vanishing soft limit, and at $z=z_{I}$ for each factorization channel $P_I^2(z_I) = 0$. Here $P_I$ is the sum of momenta of particles belonging to a subset $I$ such that one or two of the deformed particles belong to it. Due to the large-$z$ scaling, there is no simple pole at infinity. Performing the contour integral,
\bes
\label{eq:recursion}
A^{\mathrm{NLSM}}_n(\mathbb{I}_n) &=& - \mathrm{Res}_{z=1} \frac{\hat{A}^{\mathrm{NLSM}}_n(\mathbb{I}_n; z)}{(1-z)^2} - \sum_{I} \mathrm{Res}_{z=z_I} \frac{\hat{A}^{\mathrm{NLSM}}_n(\mathbb{I}_n; z)}{z(1-z)^2}\tr
&=& \sum_{a=2}^{n-2} s_{an} \hat{A}^{\mathrm{NLSM}\oplus\phi^3}_{n-1} (\mathbb{I}_{n-1} | n-1, a, 1; 1)\tr
&&- \sum_{I} \hat{A}_k^{\mathrm{NLSM}}(1,\dots,k; z_I)\, \frac{1}{P_I^2 (1-z_I)^2}\, \hat{A}_{n-k+2}^{\mathrm{NLSM}}(n-k+2,\dots,n; z_I).\qquad\quad
\ens
In this formula $P_I^2$ is the sum over un-deformed momenta. This defines a linear on-shell recursion relation.

In order to close the recursion relations, we also need a prescription for the mixed NLSM and biadjoint scalar amplitudes. In this case we use the 3-shift $(ijn)$ with $i,j$ biadjoint scalars and $n$ a sigma particle. In this case we can consider the simpler rational function
\be
\frac{\hat{A}^{\mathrm{NLSM}\oplus\phi^3}_n(\alpha|\beta; z)}{(1-z)}
\en
which vanishes for large values of $z$, since the amplitude scales as $\mathcal{O}(z^0)$, as shown in Appendix \ref{sec:large z}.

Considering again the identity
\be
\hat{A}^{\mathrm{NLSM}\oplus\phi^3}_n(\alpha|\beta) = -\oint_{|z|=\epsilon} dz\; \frac{\hat{A}^{\mathrm{NLSM}\oplus \phi^3}_n(\alpha|\beta; z)}{z(1-z)}
\en
we obtain a similar result to (\ref{eq:recursion}). Additionally, there is no pole at $z=1$, so we need to sum over the factorization channels only. Each factorized amplitude contains at most three biadjoint scalars, so the algorithm is closed.

Finally, it is worth mentioning that the only seed amplitude needed for the construction of the whole extension of the NLSM is the three-point biadjoint scalar amplitude! Just as in Yang-Mills and gravity, the three-particle amplitude becomes the fundamental building block. The amplitude of interest now is quite trivial and simply given by
\be
A^{\phi^3}_3 (\alpha | \beta) = \mathrm{sgn}(\alpha|\beta),
\en
i.e., $+1$ if $\alpha$ and $\beta$ are the same up to cyclic permutation and $-1$ otherwise.

\subsection{Examples}

Let us present a few practical examples of the recursion relations introduced above. We compute both pure NLSM and mixed amplitudes with NLSM scalars denoted by $\Sigma$ and biadjoint scalars denoted by $\phi$. In the following examples, we study the amplitudes in which the $U(\tilde{N})$ partial ordering for $\phi$ particles is the one obtained by ignoring the $\Sigma$ labels in the $U(N)$ ordering, e.g., $A_5 (1^\phi, 2^\phi, 3^\Sigma, 4^\phi, 5^\Sigma)$ denotes $A^{\mathrm{NLSM}\oplus\phi^3}_5 (1,2,3,4,5\, |\, 1,2,4)$.

\subsubsection*{4-point Amplitudes}

Knowing that all odd-multiplicity amplitudes for pure NLSM theory vanish, there are no factorization channels in the 4-point amplitude. Only the terms coming from the pole at $z=1$ contribute. Performing a $(124)$ 3-shift and using the form of the residue at $z=1$:
\be
\label{eq:ssss ampl}
A_4 (1^\Sigma, 2^\Sigma, 3^\Sigma, 4^\Sigma) = -s_{24}\,A_3^{\phi^3}.
\en
but the three particle amplitude, $A_3^{\phi^3}$, in the cubic biadjoint scalar theory is one. This produces the well-known result
$A_4 (1^\Sigma, 2^\Sigma, 3^\Sigma, 4^\Sigma) = -s_{24}$.

Due to the property of the CHY integral mentioned above, this is also equal to the case of mixed amplitudes:
\be
\label{eq:sspp ampl}
A_4 (1^\Sigma, 2^\Sigma, 3^\phi, 4^\phi) = A_4 (1^\Sigma, 2^\phi, 3^\Sigma, 4^\phi) = -s_{24}.
\en

\subsubsection*{5-point Amplitudes}

Pure NLSM 5-point amplitudes vanish. Let us demonstrate how the recursion relations work when starting with a mixed theory instead. Performing a $(125)$ 3-shift we have:
\bes
\label{eq:pppss ampl}
A_5 (1^\phi, 2^\phi, 3^\phi, 4^\Sigma, 5^\Sigma) &=& \hat{A}_3 (\hat{1}^\phi, \hat{2}^\phi, {\hat{P}_{12}}^\phi)\,  \frac{1}{(1-z)s_{\hat{1}\hat{2}}}\, \hat{A}_4 ({\hat{P}_{12}}^\phi, 3^\phi, 4^\Sigma, \hat{5}^\Sigma) \Big|_{z=z_{12}}\tr
&&\qquad+\; \hat{A}_3(\hat{2}^\phi, 3^\phi, {\hat{P}_{23}}^\phi)\, \frac{1}{(1-z)s_{\hat{2}3}}\, \hat{A}_4 ({\hat{P}_{23}}^\phi, 4^\Sigma, \hat{5}^\Sigma, \hat{1}^\phi)\Big|_{z=z_{23}}\tr
&=&\; \frac{s_{34} + s_{4\hat{5}}}{(1-z)s_{\hat{1}\hat{2}}}\Bigg|_{z=z_{12}} +\; \frac{s_{4\hat{5}} + s_{\hat{5}\hat{1}}}{(1-z)s_{\hat{2}3}}\Bigg|_{z=z_{23}},
\ens
where we have used the results (\ref{eq:sspp ampl}) in the second equality. We can compute the first factor using the following trick:
\be
\label{eq:trick}
0=\oint \frac{dz}{z(1-z)} \frac{s_{34} + s_{4\hat{5}}}{s_{\hat{1}\hat{2}}} = \frac{s_{34} + s_{45}}{s_{12}} - \frac{s_{34} + s_{4\hat{5}}}{s_{\hat{1}\hat{2}}}\Bigg|_{z=1} -\; \frac{s_{34} + s_{4\hat{5}}}{(1-z)s_{\hat{1}\hat{2}}}\Bigg|_{z=z_{12}},
\en
but since $k_5$ vanishes at $z=1$, we have $s_{34} = s_{\hat{1}\hat{2}} |_{z=1}$. Thus
\be
\frac{s_{34} + s_{4\hat{5}}}{(1-z)s_{\hat{1}\hat{2}}}\Bigg|_{z=z_{12}} = \frac{s_{34} + s_{45}}{s_{12}} -1.
\en
Repeating the same procedure for the second term in (\ref{eq:pppss ampl}) yields the final answer:
\be
\label{eq:pppss ampl result}
A_5 (1^\phi, 2^\phi, 3^\phi, 4^\Sigma, 5^\Sigma) = \frac{s_{34} + s_{45}}{s_{12}} + \frac{s_{45} + s_{15}}{s_{23}} - 1.
\en
In a similar fashion we can obtain the result:
\be
\label{eq:ppsps ampl}
A_5 (1^\phi, 2^\phi, 3^\Sigma, 4^\phi, 5^\Sigma) = \frac{s_{34} + s_{45}}{s_{12}} - 1.
\en
These results have also been checked against the corresponding CHY formulas.

\subsubsection*{6-point Amplitudes}

The computation of a pure NLSM 6-point amplitude involves both types of residues. Using a $(156)$ 3-shift:
\bes
\label{eq:NLSM 6-pt}
A_6 (1^\Sigma, 2^\Sigma, 3^\Sigma, 4^\Sigma, 5^\Sigma, 6^\Sigma) &=& - s_{26} A_5 (1^\phi,2^\phi,3^\Sigma,4^\Sigma,5^\phi) - s_{36} A_5 (1^\phi,2^\Sigma,3^\phi,4^\Sigma,5^\phi)\tr
&& - s_{46} A_5 (1^\phi,2^\Sigma,3^\Sigma,4^\phi,5^\phi)\tr
&& + \hat{A}_4 (\hat{1}^\Sigma, 2^\Sigma, 3^\Sigma, {\hat{P}_{123}}^\Sigma) \frac{1}{(1-z)^2 s_{123}} \hat{A}_4 ({\hat{P}_{123}}^\Sigma, 4^\Sigma, \hat{5}^\Sigma, \hat{6}^\Sigma)\Big|_{z=z_{123}} \tr
&& + \hat{A}_4 (\hat{6}^\Sigma, \hat{1}^\Sigma, 2^\Sigma, {\hat{P}_{345}}^\Sigma)\frac{1}{(1-z)^2 s_{345}} \hat{A}_4 ({\hat{P}_{345}}^\Sigma, 3^\Sigma, 4^\Sigma, \hat{5}^\Sigma)\Big|_{z=z_{345}} \tr
&=& \frac{1}{2} \frac{(s_{12} + s_{23})(s_{45} + s_{56})}{s_{123}} - s_{12} + \mathrm{cycl},
\ens
where we have used the on-shell amplitudes (\ref{eq:ssss ampl}, \ref{eq:pppss ampl result}, \ref{eq:ppsps ampl}). Again, replacement of any two NLSM particles with two biadjoint scalars yields the same result:
\be
A_6 (1^\Sigma, 2^\Sigma, 3^\phi, 4^\Sigma, 5^\phi, 6^\Sigma) = \frac{1}{2} \frac{(s_{12} + s_{23})(s_{45} + s_{56})}{s_{123}} - s_{12} + \mathrm{cycl}.
\en

\subsubsection*{Higher-point Amplitudes}

Here we just present the results for some higher-point amplitudes that can easily be confirmed using the new recursion relations as well as their corresponding CHY representations:
\bes
&&A_7 (1^\phi, 2^\phi, 3^\phi, 4^\Sigma, 5^\Sigma, 6^\Sigma, 7^\Sigma) = \frac{1}{s_{12}}\Bigg(\frac{(s_{34}+s_{45})(s_{67}+s_{712})}{s_{345}}+\frac{(s_{45}+s_{56})(s_{712}+s_{123})}{s_{456}}\tr
&&\qquad+\frac{(s_{56}+s_{67})(s_{123}+s_{34})}{s_{567}}-s_{34}-s_{45}-s_{56}-s_{67}-s_{712}-s_{123}\Bigg)\tr
&&\qquad+\frac{1}{s_{23}}\Bigg(\frac{(s_{45}+s_{56})(s_{71}+s_{123})}{s_{456}}+\frac{(s_{56}+s_{67})(s_{123}+s_{234})}{s_{567}}+\frac{(s_{67}+s_{71})(s_{234}+s_{45})}{s_{671}}\tr
&&\qquad-s_{45}-s_{56}-s_{67}-s_{71}-s_{123}-s_{234}\Bigg) +\frac{(s_{67}+s_{71})(s_{34}+s_{45})}{s_{671}s_{345}}\tr
&&\qquad-\frac{s_{34}+s_{45}}{s_{345}}-\frac{s_{45}+s_{56}}{s_{456}}-\frac{s_{56}+s_{67}}{s_{567}}-\frac{s_{67}+s_{71}}{s_{671}}+2.
\ens
We also computed the 8-pt and 10-pt pure NLSM amplitudes and confirmed they agree with the ones presented in \cite{Kampf:2013vha}, obtained using Berends-Giele recursions relations.

\section{\label{sec:Galileon}Extension of the Special Galileon Theory}

Galileon theories arise as effective field theories in the decoupling limit of massive gravity \cite{Hinterbichler:2011tt} and the Dvali-Gabadadze-Poratti model \cite{Dvali:2000hr}. Scattering amplitudes in Galileon theories were studied in \cite{Kampf:2014rka}. These Lorentz invariant theories of scalars are named after the internal Galilean symmetry they possess.

A special class of Galileon theories with soft limits that vanish particularly fast was discovered in \cite{Cheung:2014dqa, CHY}. The underlying additional symmetry likely responsible for this behavior was identified in \cite{Hinterbichler:2015pqa}.

The CHY representation for this special Galileon theory was proposed in \cite{CHY} and it is given by
\be
A_n^{\rm sGal:CHY} = \oint d\mu_n (\pf' \A_n)^4.
\en
In this section we study the extension of the special Galileon making use of the Kawai-Lewellen-Tye (KLT) relations \cite{KLT} and the results obtained in the previous section for the NLSM.

\subsection{Single Soft Limits using KLT Relations}

CHY formulas provide a simple way of understanding the KLT relations. It was shown in \cite{Cachazo:2013iea} that the KLT kernel matrix is the inverse of a matrix whose entries are biadjoint $\phi^3$ scalar amplitudes. In order to state the general version of the relations it is important to start with a theory in CHY form, where the integrand has been decomposed into two half integrands $\mathcal{I}_L$ and $ \mathcal{I}_R$. This decomposition is not arbitrary. Each half-integrand must carry exactly half the $\mathrm{SL}(2,\mathbb{C})$ weight as the full integrand and hence their name. Assuming that the ``target" theory we want to construct using KLT is given by
\be
A^{\rm target}_n = \oint d\mu_n \; \mathcal{I}_L \; \mathcal{I}_R
\en
then
\be
A^{\rm target}_n = \sum_{\omega,\tilde{\omega} \in S_{n-3}} \left( \oint d\mu_n \; \mathcal{I}_L \; \mathcal{C}_n(\omega) \right) \left[ A^{\phi^3}_n (\omega | \tilde{\omega}) \right]^{-1} \left( \oint d\mu_n \; \mathcal{C}_n(\tilde{\omega}) \; \mathcal{I}_R \right).
\en
Therefore mixing and matching the factors $\mathcal{I}_L$, $\mathcal{I}_R$ produces a zoo of relations among amplitudes in different theories.

Some examples of interest for this work are: special Galileon $=$ NLSM $\overset{\mathrm{\scriptscriptstyle KLT}}{\otimes}$ NLSM, and Born-Infeld $=$ NLSM $\overset{\mathrm{\scriptscriptstyle KLT}}{\otimes}$ Yang-Mills.

Note that the ``left'' and ``right'' theories in the KLT formula are both ordered with respect to at least one group factor. The sum in the KLT formula proceeds over any subsets with $(n-3)!$ elements out of the $n!$ possible permutations of $n$ elements such that the determinant of  $A^{\phi^3}_n (\omega | \tilde{\omega})$ is non zero. One common choice of valid permutations is obtained by fixing three labels and permuting the remaining $n-3$.

KLT relations have been used to calculate soft limits in \cite{BjerrumBohr:2010hn, Du:2014eca}. Here we give a simple derivation of this result using the above interpretation of the KLT kernel. In particular, the explicit form of the KLT matrix is not needed! Choosing the following basis and taking the particle $n$ to be soft we obtain, 
\be
A^{\phi^3}_n( 1,\omega,n-1,a,n | 1,\tilde{\omega},n-1,n,b) = \tau^{-1} \frac{\delta_{ab}}{\hat{s}_{an}} A^{\phi^3}_{n-1} (1,\omega,n-1,a | 1,\tilde{\omega},n-1,a) + \mathcal{O}(1),\nonumber
\en
since this is the only factorization channel contributing to the leading order. The sets $\omega$, $\tilde{\omega}$ now consist of $n-4$ particles. If we label the rows and columns by $a$ and $b$ respectively, the matrix becomes block diagonal in $a,b$ in the soft limit. We can exploit this fact to directly obtain inverses of each block from the inverse of the whole matrix.

We are now fully equipped to study the soft limits of a given theory knowing only the soft limits of its KLT factors. Before proceeding with the Galileon theory, let us illustrate the technique with Einstein gravity written as a square of Yang-Mills:
\be
A^{\mathrm{GR}}_n = \sum_{a,b = 2}^{n-2}\; \sum_{\omega,\tilde{\omega} \in S_{n-4}} A^{\mathrm{YM}}_n(1,\omega,n-1,a,n) \left[ A^{\phi^3}_n \right]^{-1} A^{\mathrm{YM}}_n(1,\tilde\omega,n-1,n,b),
\en
where for brevity of notation we leave the ordering of the kernel matrix implicit. Using the familiar soft factor for Yang-Mills theory we obtain in the soft limit,
\bes
A^{\mathrm{GR}}_n &\to& \tau^{-1} \sum_{a=2}^{n-2} \hat{s}_{an} \left( \frac{\epsilon_n \cdot k_a}{\hat{k}_n \cdot k_a} - \frac{\epsilon_n \cdot k_1}{\hat{k}_n \cdot k_1} \right) \left( \frac{\tilde{\epsilon}_n \cdot k_{n-1}}{\hat{k}_n \cdot k_{n-1}} - \frac{\tilde{\epsilon}_n \cdot k_a}{\hat{k}_n \cdot k_a} \right) \left( A_{n-1}^{\mathrm{YM}} \overset{\mathrm{\scriptscriptstyle KLT}}{\otimes} A_{n-1}^{\mathrm{YM}} \right)\tr
&=& - \tau^{-1} \left( \sum_{a=1}^{n-1} \frac{\epsilon_{\mu\nu} k_a^{\mu} k_a^\nu}{\hat{k}_n \cdot k_a} \right) A^{\mathrm{GR}}_{n-1},
\ens
where we have used momentum conservation to obtain what is the correct soft factor for Einstein gravity.

We can now focus on the theory under consideration, the special Galileon:
\be
A^{\mathrm{sGal}}_n = \sum_{a,b = 2}^{n-2}\; \sum_{\omega,\tilde{\omega} \in S_{n-4}} A^{\mathrm{NLSM}}_n(1,\omega,n-1,a,n) \left[ A^{\phi^3}_n \right]^{-1} A^{\mathrm{NLSM}}_n(1,\tilde\omega,n-1,n,b),
\en
where for brevity of notation we leave the ordering of the kernel matrix implicit. Using (\ref{eq:NLSM extension}), in the soft limit the amplitude becomes
\bes
A^{\mathrm{sGal}}_n &\to& \tau^3\, \sum_{a=2}^{n-2}\; \sum_{\omega,\tilde{\omega} \in S_{n-4}} \left( \sum_{\substack{c=2 \\ c \neq a}}^{n-1} \hat{s}_{cn} A_{n-1}^{\mathrm{NLSM} \oplus \phi^3} (a,c,1 | 1, \omega, n-1, a)\right) \hat{s}_{an} \left[A_{n-1}^{\phi^3}\right]^{-1} \tr
&&\quad\quad\quad\quad\quad\quad\quad\quad \times\left( \sum_{\substack{d=1 \\ d \neq a}}^{n-2} \hat{s}_{dn} A_{n-1}^{\mathrm{NLSM} \oplus \phi^3} (1,\tilde{\omega},n-1,a | n-1,d,a)\right) +\mathcal{O}(\tau^4)\tr
&=& \tau^3\, \sum_{a=2}^{n-2} \sum_{\substack{c=2 \\ c \neq a}}^{n-1}  \sum_{\substack{d=1 \\ d \neq a}}^{n-2} \hat{s}_{an} \hat{s}_{cn} \hat{s}_{dn} \left( A_{n-1}^{\mathrm{NLSM} \oplus \phi^3} (a,c,1 | \cdot) \overset{\mathrm{\scriptscriptstyle KLT}}{\otimes} A_{n-1}^{\mathrm{NLSM} \oplus \phi^3} (\cdot | n-1,d,a) \right).\nonumber
\ens
The last equality holds up to terms of order $\mathcal{O}(\tau^4)$.

We can identify the theory controlling the soft limit as a composite theory of special Galileons $\pi$, $U(N)$ NLSM bosons $\Sigma$, $U(\tilde N)$ NLSM bosons $\tilde{\Sigma}$, and $U(N) \times U(\tilde N)$ biadjoint scalars $\phi$. Written as a CHY integral it gives us,
\be
\label{eq:composite}
A_{n}^{\mathrm{sGal}\oplus\mathrm{NLSM}^2\oplus\phi^3}(\alpha | \beta) = \oint d\mu_n \; \Big( \mathcal{C}(\alpha)\, (\pf\, \A_{\overbar{\alpha}})^2 \Big) \Big( \mathcal{C}(\beta)\, (\pf\, \A_{\overbar{\beta}})^2 \Big).
\en
This is a natural generalization of the result (\ref{eq:NLSM-phi}). We used similar notation, i.e., the particles with labels belonging to $\alpha \cap \beta$ are biadjoint scalars, those in $\alpha \cap \overbar{\beta}$ and $\overbar{\alpha} \cap \beta$ are $U(N)$ and $U(\tilde N)$ NLSM particles respectively, and finally those with labels $\overbar{\alpha} \cap \overbar{\beta}$ are the special Galileons.

We have therefore identified the extension of the special Galileon theory which governs its single soft limit. The final form of the limit is then
\be
\boxed{A^{\mathrm{sGal}}_n = \tau^3\, \sum_{a=2}^{n-2} \sum_{\substack{c=2 \\ c \neq a}}^{n-1}  \sum_{\substack{d=1 \\ d \neq a}}^{n-2} \hat{s}_{an} \hat{s}_{cn} \hat{s}_{dn}\, A_{n-1}^{\mathrm{sGal}\oplus\mathrm{NLSM}^2\oplus\phi^3}(a,c,1 | n-1,d,a) +\mathcal{O}(\tau^4).}
\en

\subsection{\label{sec:composite}Examples}

The integral (\ref{eq:composite}) defines a mixed theory of multiple interacting species of scalars. Here we provide several low-point examples of the scattering amplitudes. For $n=4$ we have
\be
A_4(1^\pi,2^\pi,3^\Sigma,4^\Sigma)=-s_{12}\, s_{13}\, s_{14}
\en
which, as in the previous section, agrees with $A_4(1^\pi,2^\pi,3^\pi,4^\pi)$.

For $n=5$ we have the examples,
\be
A_5(1^{\tilde{\Sigma}}, 2^{\tilde{\Sigma}}, 3^\Sigma, 4^\Sigma, 5^\phi) = s_{34} (s_{12} + s_{23} - s_{45}) + s_{12} (s_{23} - s_{15}) + s_{15}s_{45},
\en
\bes
A_5(1^\pi, 2^\pi, 3^\phi, 4^\phi, 5^\phi) &=& s_{12}s_{23}\left(\frac{s_{51}+s_{12}}{s_{34}}+\frac{s_{12}+s_{23}}{s_{45}}-1\right) \tr
&&+ s_{12}s_{24}\left(\frac{s_{51}+s_{12}}{s_{34}}+\frac{s_{12}+s_{24}}{s_{35}}-1\right),
\ens
\bes
A_5(1^\pi, 2^{\tilde{\Sigma}}, 3^\Sigma, 4^\phi, 5^\phi) &=& s_{23}s_{45}\left(\frac{s_{51}+s_{12}}{s_{34}}+\frac{s_{12}+s_{23}}{s_{45}}-1\right)\left(\frac{s_{12}}{s_{45}} + \frac{s_{34}}{s_{25}} -1 \right)\tr
&&+s_{24}s_{35}\left(-\frac{s_{51}+s_{12}}{s_{34}}-\frac{s_{12}+s_{24}}{s_{35}}+1\right)\left( \frac{s_{34}}{s_{25}} + \frac{s_{15}}{s_{24}} -1 \right).\qquad\quad
\ens

Our last example is actually an amplitude that lives purely in the subsector of the extended Galileon theory that coincides with the extended NLSM discussed in the previous section:
\bes
A_6 (1^\Sigma, 2^\Sigma, 3^\phi, 4^\phi, 5^\phi, 6^\phi) &=&\frac{1}{s_{56}}\left(-\frac{s_{12}+s_{23}}{s_{123}}+1\right)+\frac{1}{s_{34}}\left(-\frac{s_{21}+s_{16}}{s_{612}}+1\right) \tr
&-& \frac{s_{12}+s_{234}}{s_{34}s_{56}} + \frac{1}{s_{45}}\left(-\frac{s_{12}+s_{23}}{s_{123}}-\frac{s_{21}+s_{16}}{s_{612}}+1\right).\quad
\ens
The reason it is more natural to present this amplitude here instead of the previous section is that it was computed purely from the CHY representation and not from the BCFW-like recursion relation.

Once again, notice that the CHY formula makes it clear that some amplitudes are equal to each other. For instance, replacing two Galileons in a pure Galileon amplitude with two sigma particles in the same flavor group gives the same result.

Finally, let us mention that while we now have control over the single soft limit of the special Galileon theory this is not yet enough to allow the use of a simple recursion relation as the one presented for the NLSM in the previous section. It would be interesting to explore this issue further.

\section{\label{sec:BI}Extension of Born-Infeld Theory}

Born-Infeld theory is a non-linear generalization of Maxwell theory still based on an abelian group. Its CHY representation is given by \cite{CHY}
\be
A_n^{\rm BI:CHY} = \oint d\mu_n (\pf' \A_n)^2~\pf'\Psi_n.
\en
where $\Psi_n$ is a $2n \times 2n$ antisymmetric matrix which contains all the information regarding the polarization vectors $\epsilon_a^\mu$ of the $n$ BI photons. The precise form of the matrix is not used here, but it is given in the review presented in Appendix \ref{sec:Review CHY}.

The form of the CHY representation suggests that the derivation of the soft limits should proceed in the same way as in the Galileon case by using the KLT procedure. In this case, the BI theory is constructed by using one copy of the NLSM and one copy of a $U(N)$ Yang-Mills gauge theory.

Repeating the steps of the previous section we obtain
\bes
A^{\mathrm{BI}}_n &=& \sum_{a,b = 2}^{n-2}\; \sum_{\omega,\tilde{\omega} \in S_{n-4}} A^{\mathrm{NLSM}}_n(1,\omega,n-1,a,n) \left[ A^{\phi^3}_n \right]^{-1} A^{\mathrm{YM}}_n(1,\tilde\omega,n-1,n,b) \tr
&\to& \tau\, \sum_{a=2}^{n-2} \sum_{\substack{c=2 \\ c \neq a}}^{n-1} \hat{s}_{an} \hat{s}_{cn} \mathcal{\hat{S}}^{\mathrm{YM}}(n-1,n,a) \left( A_{n-1}^{\mathrm{NLSM} \oplus \phi^3} (a,c,1 | \cdot) \overset{\mathrm{\scriptscriptstyle KLT}}{\otimes} A_{n-1}^{\mathrm{YM}} (\cdot) \right) +\mathcal{O}(\tau^2),\nonumber
\ens
where the standard Yang-Mills soft factor for particle $n$ is given by
\be
\hat{\mathcal{S}}^{\mathrm{YM}}(a,n,b) = \frac{\epsilon_n \cdot k_a}{\hat{k}_n \cdot k_a} - \frac{\epsilon_n \cdot k_b}{\hat{k}_n \cdot k_b}.
\en
We can again write the term in the brackets using a CHY representation,
\be
A_{n}^{\mathrm{BI}\oplus\mathrm{YM}}(\alpha) = \oint d\mu_n \; \Big( \mathcal{C} (\alpha)\, (\pf\, \A_{\overbar{\alpha}})^2 \Big) \Big( \pf' \Psi_n \Big).
\en
We interpret these CHY formulas as computing amplitudes in a theory including gluons and Born-Infeld photons as external particles.

We are now ready to combine all the results to obtain an expression for the single soft limit of the Born-Infeld theory as a linear combination of amplitudes in its extension:
\be
\boxed{A^{\mathrm{BI}}_n = -\tau\, \sum_{a=2}^{n-2} \sum_{\substack{c=2 \\ c \neq a}}^{n-1} \hat{s}_{an} \hat{s}_{cn} \left( \frac{\epsilon_n \cdot k_{n-1}}{\hat{k}_n \cdot k_{n-1}} - \frac{\epsilon_n \cdot k_a}{\hat{k}_n \cdot k_a} \right)  A_{n-1}^{\mathrm{BI}\oplus \mathrm{YM}}(a,c,1) +\mathcal{O}(\tau^2).}
\en

We end this section by providing some illustrative examples. In order to present compact formulas we work in four dimensions. Denoting Born-Infeld photons as $\gamma$ and gluons as $g$, we find
\be
A^{\mathrm{BI}\oplus\mathrm{YM}}_4 (1^{\gamma^-}, 2^{\gamma^-}, 3^{g^+}, 4^{g^+}) = -s_{12}(s_{12}+s_{13}) \frac{\la 12 \ra^3}{\la 23 \ra \la 34 \ra \la 41 \ra},
\en
and
\bes
A^{\mathrm{BI}\oplus\mathrm{YM}}_5 (1^{\gamma^-}, 2^{\gamma^-}, 3^{g^-}, 4^{g^+}, 5^{g^+}) &=& s_{23}\left(s_{45}-s_{13}-s_{23}\right) \frac{[45]^3}{[12][23][34][51]}\tr
&&+ s_{24}\left(s_{14}+s_{24}-s_{35}\right) \frac{[45]^4}{[12][24][43][35][51]}.\qquad
\ens

\section{\label{sec:DBI-VA}Extension of DBI-Volkov-Akulov Theory}

The Dirac-Born-Infeld-Volkov-Akulov (DBI-VA) theory \cite{Volkov:1973ix, Kallosh:1997aw, Bergshoeff:2013pia} is the fully supersymmetric effective theory on a D-brane. A D-brane not only spontaneously breaks part of the Poincar\'e group, but also half of the supersymmetries.
In this section we consider the maximally supersymmetric DBI-VA theory in four dimensions. The theory linearly realizes $16$ supercharges while $16$ others are non-linearly realized. Here we present the explicit CHY, or Witten-RSV, formula for the complete S-matrix of the theory. It is interesting to note that the full action for the theory was not known until 2013 when Bergshoeff et.al. gave the explicit form in \cite{Bergshoeff:2013pia}.

When the external particles are restricted to be fermions (Goldstinos) with R-charge indices of only two kinds, e.g., $(123)$ and $(4)$, the theory is simply known as the Volkov-Akulov theory and recursion relations for its amplitudes have been recently studied in \cite{Chen:2014xoa, Luo:2015tat}.

Before proceeding to the DBI-VA theory, let us introduce some notions that allow the transition from arbitrary dimensions to four dimensions, and hence from CHY formulas to Witten-RSV formulas \cite{Witten, RSV}.

In four dimensions, it is most convenient to introduce a set of spinor-valued polynomials in $\sigma$ of degree $d$ and $\tilde{d}$ respectively, where $d+\tilde{d} = n-2$ and $d$ taking values in $\{ 1,2,\dots, n-3\}$,
\be
\lambda_\alpha^{(d)}(\sigma) = \sum_{i=1}^{d} \rho_\alpha^{(i)} \sigma^i, \qquad \tilde{\lambda}_{\dot{\alpha}}^{(\tilde{d})}(\sigma) = \sum_{i=1}^{\tilde{d}} \tilde{\rho}_\alpha^{(i)} \sigma^i.
\en
It turns out that the scattering equations are equivalent to imposing that
\be
(\lambda_a)_\alpha (\tilde{\lambda}_a)_{\dot{\alpha}} = \oint_{|\sigma-\sigma_a|=\epsilon} \frac{\lambda_\alpha^{(d)}(\sigma) \tilde{\lambda}_{\dot{\alpha}}^{(\tilde{d})}(\sigma)}{\prod_{b=1}^{n} (\sigma - \sigma_b)}
\en
for all particles \cite{Cachazo:2013iaa}.

In the case of gravity and Yang-Mills, it is known that the equations with $d=k-1$ possess $\eulerian{n-3}{k-2}$ solutions (triangular brackets denote the Eulerian numbers \cite{Eulerian}). Gravity and Yang-Mills amplitudes in the $\mathrm{N^{k-2}MHV}$ sector can be constructed using only solutions for $d=k-1$. Of course, the complete supersymmetric amplitude obtained by summing over all sectors requires the use of the usual number of solutions, $\sum_{k=2}^{n-2} \eulerian{n-3}{k-2} = (n-3)!$.

CHY integral representations become Witten-RSV formulas as follows
\be
A_{n,k} = \oint d\Omega^{\mathrm{B}}_{n,k} \; \mathcal{I}_L (\{k, \epsilon, \sigma, \rho,\tilde\rho, t,\tilde{t} \}) \; \mathcal{I}_R (\{k, \tilde\epsilon, \sigma, \rho, \tilde\rho, t, \tilde{t} \}),
\en
with the bosonic measure defined as
\bes
d\Omega^{\mathrm{B}}_{n,k} &\equiv& \frac{1}{\mathrm{vol\, SL}(2,\mathbb{C}) \times \mathrm{GL}(1,\mathbb{C})} \prod_{a=1}^n d\sigma_a\, dt_a\, d\tilde{t}_a\, \delta \left( t_a \tilde{t}_a - \frac{1}{\prod_{b \neq a} \sigma_{ab} } \right) \tr
&& \times \prod_{i=0}^{d} d^2\rho^{(i)}\, \prod_{a=1}^n \delta^2 \big( \lambda_a - t_a \lambda(\sigma_a) \big) \prod_{i=0}^{\tilde{d}} d^2\tilde{\rho}^{(i)}\, \prod_{a=1}^n \delta^2 \big( \tilde{\lambda}_a - \tilde{t}_a \tilde{\lambda}(\sigma_a) \big)\, \frac{1}{R(\lambda) R(\tilde{\lambda})}.\nonumber
\ens
The additional variables $t_a, \tilde{t}_a$ are introduced to fix the little group scaling. $R(\lambda)$ denotes the resultant of the polynomials $\lambda_1^{(d)} (\sigma), \lambda_2^{(d)} (\sigma)$. A similar definition holds for $R(\tilde{\lambda})$.

Half-integrands that can be used to describe supersymmetric theories are easy to construct. More concretely, for $\mathcal{N}=4$ super Yang-Mills with partial ordering $\alpha$,
\be
\mathcal{I}_L = \mathcal{C}_n(\alpha), \qquad \mathcal{I}_R = \oint d^2 \Omega_{n,k}^{\mathrm{F}}\, R(\lambda) R(\tilde{\lambda}),
\en
and for $\mathcal{N}=8$ supergravity
\be
\mathcal{I}_L = \oint d^2 \Omega_{n,k}^{\mathrm{F}}\, R(\lambda) R(\tilde{\lambda}), \qquad \mathcal{I}_R = \oint d^2 \tilde{\Omega}_{n,k}^{\mathrm{F}}\, R(\lambda) R(\tilde{\lambda}).
\en
Here $d^2 \Omega_{n,k}^{\mathrm{F}}$ denotes a measure over fermionic maps whose exact form can be found in \cite{Cachazo:2013zc}.

Let us further expand the dictionary of supersymmetric theories in this formalism. We first recall that the factor $\pf' \A_n$ vanishes identically when evaluated on solutions other than the middle charge sector, $d=\tilde{d}=n/2-1$ \cite{CHY}. For this helicity conserving solution we have
\be
\pf' \A_n \big|_{k=\frac{n}{2}} = \frac{R(\lambda) R(\tilde{\lambda})}{|1 2 \cdots n|}.
\en
The term in the denominator is the Vandermonde determinant, $|1 2 \cdots n| = \prod_{a<b}\sigma_{ba}$. We can now define the superamplitude in a maximally supersymmetric generalization of Born-Infeld theory as
\be
\mathcal{A}^{\mathrm{sBI}}_n = \oint d\Omega^{\mathrm{B}}_{n,\frac{n}{2}} \oint d^2 \Omega_{n,\frac{n}{2}}^{\mathrm{F}}\; |1 2 \cdots n|\, (\pf' \A_n)^3.
\en
We propose that this formula computes the complete tree-level S-matrix of the maximally supersymmetric DBI-VA theory.

A simple consequence of the previous construction is that these amplitudes can be computed using a KLT product of NLSM and  $\mathcal{N}=4$ super Yang-Mills amplitudes:
\be
A^{\mathrm{DBI-VA}}_n = A^{\mathrm{NLSM}}_n \overset{\mathrm{\scriptscriptstyle KLT}}{\otimes} A_n^{\mathcal{N}=4\; \mathrm{SYM}}.
\en

Some simple checks can be ontained by restricting the ${\cal N}=4$ SYM amplitude to only external fermions. Here we denote fermions with R-charge $(123)$ by $\psi$ and those with $(4)$ by $\bar\psi$. With this definition we can compute the first two non-vanishing amplitudes,
\be
A_4^{\mathrm{DBI-VA}}(1^\psi, 2^{\bar{\psi}}, 3^\psi, 4^{\bar{\psi}}) = s_{13} \la 24 \ra [13],
\en
and $A_6^{\mathrm{DBI-VA}}(1^\psi, 2^{\bar{\psi}}, 3^\psi, 4^{\bar{\psi}}, 5^\psi, 6^{\bar{\psi}})$. The latter was recently computed using a recursion relation exploiting the soft limits in \cite{Luo:2015tat}, where a compact formula was given:
\begin{align}
A_6^{\mathrm{DBI-VA}}(1^\psi, 2^{\bar{\psi}}, 3^\psi, 4^{\bar{\psi}}, 5^\psi, 6^{\bar{\psi}}) =
&&\frac{s_{15}s_{26}}{s_{145}} \la 26\ra [15] \la 4|1+5|3] + \frac{s_{13}s_{46}}{s_{123}} \la 46 \ra [13] \la 2 | 1+3 |5]\tr\displaybreak
&&+ \frac{s_{35}s_{24}}{s_{356}} \la 24 \ra [35] \la 6 |3+5| 1] + \frac{s_{15}s_{46}}{s_{125}} \la 64 \ra [15] \la 2 | 1+5 | 3]\tr
&&+ \frac{s_{35}s_{26}}{s_{345}} \la 62 \ra [35] \la 4 | 3+5 |1] + \frac{s_{13}s_{26}}{s_{134}} \la 62 \ra [13] \la 4 |1+3| 5] \tr
&&+ \frac{s_{13}s_{24}}{s_{136}} \la 24 \ra [13] \la 6 | 1+3| 5] + \frac{s_{35}s_{46}}{s_{235}} \la 64 \ra [53] \la 2 | 3+5 | 1] \tr
&&+ \frac{s_{15}s_{24}}{s_{156}} \la 42 \ra [15] \la 6 | 1+5 | 3].\nonumber
\end{align}
We find that our results perfectly match the expressions in \cite{Luo:2015tat}.

Finally, we can study the single soft limits of the DBI-VA theory and discover its extension. A derivation analogous to the one from Section \ref{sec:BI} yields:
\be
\boxed{A^{\mathrm{DBI-VA}}_n = \tau\, \sum_{a=2}^{n-2} \sum_{\substack{c=2 \\ c \neq a}}^{n-1} \hat{s}_{an} \hat{s}_{cn}\, \mathcal{\hat{F}}(n) A_{n-1}^{\mathrm{DBI-VA}\,\oplus\,\mathcal{N}=4\; \mathrm{SYM}}(a,c,1) +\mathcal{O}(\tau^2),}
\en
where $\mathcal{\hat{F}}(n)$ is the soft factor of $\mathcal{N}=4$ SYM. In the case of a soft gluino this was defined in \cite{Liu:2014vva} with a factor of $\tau^{-1}$ stripped away.

\section{\label{sec:Discussion}Discussion and Outlook}

In this work we have uncovered a rich structure hiding in the single soft limit of effective field theories. While these limits vanish, the precise structure of the limit involves lower point amplitudes of a larger theory. We have only scratched the surface of this phenomenon, which we call extension via soft limits. The CHY representation of the original amplitudes makes the interpretation of the soft limit as a linear combination of the amplitudes in the extended theory manifest. While CHY formulas are a useful discovery tool, it would be interesting to find a purely effective field theory derivation of this phenomenon, along the lines of \cite{Larkoski:2014bxa}.

The extensions contain the original theory as well as additional fields and flavor/color groups. While our first example was the extension of the NLSM, it is more natural to think that the best starting point is in fact the special Galileon (sGal) theory. The extension of sGal includes two copies of NLSM and a biadjoint cubic scalar. It is well-known that both general Galileons (and hence sGal) and NLSM arise as effective theories of the St$\ddot{\rm u}$kelberg field in the decoupling limit of massive gravity and Yang-Mills theories respectively \cite{Hinterbichler:2011tt}. This immediately suggests that there is a master theory combining massive gravity, massive Yang-Mills and a biadjoint scalar which, in the decoupling limit, gives rise to the extension of the special Galileon. We leave this very interesting question for future research.

The nature of extensions involving the Born-Infeld theory are even more interesting. While the original theory only contains photons, the extended theory involves gluons as well. In order to find a Lagrangian for this theory, it would be very useful to find a D-brane realization of the extended BI theory. It is possible that our constructions could lead to formulas for the full non-linear effective theory of multiple D-branes, also known as the non-abelian DBI \cite{Tseytlin:1986ti}. We also studied supersymmetric generalizations in four dimensions of the DBI theory known as the DBI-VA theory. Here again, we find the power of on-shell methods in making large amounts of supersymmetry manifest even when there is not any known superspace formulation.

Another line of research which we only briefly started to explore is the use of extensions to ``BCFW complete" a theory \cite{Benincasa:2007xk}. While the BCFW technique is especially useful in theories with spin, its straightforward application is less so in scalar theories. The usual obstruction is related to the non-vanishing behavior of amplitudes for large values of the deformation parameter $z$. In recent years, it has become clear that the vanishing of soft limits is a resource that can be harnessed in order to tame the lack of information at infinity \cite{Cheung:2014dqa,Cheung:2015ota}. In this work we have shown how to take these ideas a step further by embedding a theory in a larger one, so that it is possible to find recursion relations that were not available for the original theory alone. We used this construction in the case of NLSM, but one should also be able to generalize this to a larger class of theories.

In recent years there has been a renewed interest in symmetries of asymptotically flat space-times. These symmetries form a group known as the BMS group, named after Bondi, van der Burg, Metzner and Sacks \cite{Bondi:1962px, Sachs:1962wk}. The resurgence of interest started in 2013 with Strominger's suggestion \cite{Strominger:2013jfa} that soft particles, such as gravitons, are Goldstone bosons of spontaneously broken symmetries. It is interesting to note that Einstein gravity was among the first theories for which a CHY representation was discovered. Most of the other theories that possess CHY representations can also be thought of as theories of Goldstone bosons, as discussed in this work. It is thus possible to suggest that CHY formulations are naturally related to a large generalization of the BMS group \cite{Amplitudes 2015}. In \cite{Amplitudes 2015} it was also proposed that there could be irreducible representations of this hypothetical group associated with each known theory of particles. Moreover, the tensor product of representations would be identified with the KLT product of theories,
\be
{\rm Theory}_3 = {\rm Theory}_1 \overset{\mathrm{\scriptscriptstyle KLT}}{\otimes} {\rm Theory}_2.
\en
It was also proposed that putting together different species of particles leads to the notion of a direct sum of representations. In this work we have taken steps towards making this notion more precise for theories that arise as extensions of other ones. In such cases the way the theories interact is built-in and the operation
\be
{\rm Theory}_3 = {\rm Theory}_1 \oplus {\rm Theory}_2
\en
is naturally defined. This is a line of research which is worth exploring especially in three dimensions, where the structure of the BMS group has been recently understood in great detail \cite{Barnich:2015uva}.

Finally, note that when CHY formulas were first introduced, only three theories were known to possess them: Einstein gravity, Yang-Mills and a biadjoint cubic scalar. The building blocks used for the construction seemed to be rigid at the time. However, further research demonstrated that new blocks are possible and a variety of theories were found to also have CHY representations. The complete space of theories that admit CHY representations is not currently known. In this work we have expanded the space even more by introducing new blocks such as minors of the matrix $\A_n$. The theories found in this work lead to a natural generalization, which includes them as particular cases,
\be
\oint d\mu_n \Big( \mathcal{C}(\alpha) \, \pf\, \textsf{X}_\beta \, \pf\, \A_\gamma \, \pf\, \A_\delta \, \pf \Psi_{\overbar{\alpha\, \cup\, \beta\, \cup\, \gamma\, \cup\, \delta}} \Big) \Big( \mathcal{C}(\mu) \, \pf\, \textsf{X}_\nu \, \pf\, \A_\rho \, \pf\, \A_\sigma \, \pf \Psi_{\overbar{\mu\, \cup\, \nu\, \cup\, \rho\, \cup\, \sigma}} \Big).\nonumber
\en
We conjecture that this simple object computes all single-trace tree-level scattering amplitudes of a theory mixing all the theories summarized in Table \ref{tab:theories} of Appendix \ref{sec:Review CHY}. Each external particle belongs to two half-integrands (or four quarter-integrands or one half-integrand and two quarter-integrand) in such a way to produce the right $\mathrm{SL}(2,\mathbb{C})$ scaling for each half-integrand separately. For instance, $a \in \alpha, \overbar{\mu\, \cup\, \nu\, \cup\, \rho\, \cup\, \sigma}$ is a gluon, $b \in \gamma, \delta, \rho, \sigma$ is a special Galileon, and $c \in \gamma, \delta, \nu, \rho$ defines a DBI scalar. As before, if a Pfaffian matrix contains all particles, it should be appropriately reduced.

\section*{Acknowledgements}

We would like to thank Song He, Zhengwen Liu and Ellis Yuan for useful comments. This work is supported by Perimeter Institute for Theoretical Physics. Research at Perimeter Institute is supported by the Government of Canada through Industry Canada and by the Province of Ontario through the Ministry of Research \& Innovation.

\renewcommand{\thefigure}{\thesection.\arabic{figure}}
\renewcommand{\thetable}{\thesection.\arabic{table}}
\appendix
\section{\label{sec:Review CHY}Review of CHY Formulas}

Cachazo-He-Yuan fomulae \cite{1307.2199 CHY,Cachazo:2013iea} provide a way of calculating the tree-level S-matrix in arbitrary dimension for a wide range of theories. Scattering amplitude for $n$ particles arises from a multidimensional contour integral over the moduli space of genus zero Riemann surfaces with $n$ punctures, $\mathcal{M}_{0,n}$. It can be summarized into the concise expression:
\be
A_n = \oint d\mu_n \; \mathcal{I}_L (\{k, \epsilon, \sigma \}) \; \mathcal{I}_R (\{k, \tilde\epsilon, \sigma\}),
\en
where $k_a, \epsilon_a, \sigma_a$ are the momentum, polarization vector, and puncture location for $a$th particle respectively. We define the measure as:

\bes
d\mu_n &\equiv& \sigma_{ij}\sigma_{jk}\sigma_{ki} \prod_{a \neq i,j,k} E_a^{-1} \; \frac{d^n \sigma}{\mathrm{vol\,SL}(2,\mathbb{C})} \tr
&=& ( \sigma_{ij}\sigma_{jk}\sigma_{ki}) (\sigma_{pq}\sigma_{qr}\sigma_{rp}) \prod_{a \neq i,j,k} E_a^{-1} \prod_{b \neq p,q,r} d\sigma_b.
\ens
$\mathrm{SL}(2,\mathbb{C})$ invariance allows to fix the positions of three punctures $\sigma_p,\sigma_q,\sigma_r$. Here $\sigma_{ab}=\sigma_a -\sigma_b$ and the scattering equations \cite{Cachazo:2013gna} are given by
\be
E_a = \sum_{b \neq a} \frac{s_{ab}}{\sigma_{ab}},
\en
where $s_{ab} = (k_a + k_b)^2 = 2 k_a \cdot k_b$ for massless momenta. They define a map from the space of kinematic invariants to $\mathcal{M}_{0,n}$ and fully localize the CHY integral on $(n-3)!$ of their solutions,
\be
A_n = \sum_{i=1}^{(n-3)!} \frac{1}{\mathrm{det}' \Phi^{(i)}_n} \; \mathcal{I}_L (\{k, \epsilon, \sigma^{(i)} \}) \; \mathcal{I}_R (\{k, \tilde\epsilon, \sigma^{(i)}\}).
\en
Here $\mathrm{det}' \Phi^{(i)}_n$ is the appropriate Jacobian factor,
\be
\mathrm{det}' \Phi^{(i)}_n = \frac{\mathrm{det}[\Phi^{(i)}_n]^{ijk}_{pqr}}{( \sigma_{ij}\sigma_{jk}\sigma_{ki}) (\sigma_{pq}\sigma_{qr}\sigma_{rp})},
\en
where we removed the rows $i,j,k$ and columns $p,q,r$ from the matrix $\Phi_n$, which for a given solution we define through
\be
[\Phi_n]_{ab} = \begin{dcases} 
	\,\frac{s_{ab}}{\sigma_{ab}^2} & a \neq b, \\
	-\sum_{c \neq a} \frac{s_{ac}}{\sigma_{ac}^2} & a = b.
\end{dcases}
\en

The theories known to have a CHY representation \cite{CHY} are summarized in Table \ref{tab:theories}.
\begin{table}[!h]
	\begin{center}
		\begin{tabular}{l|c|c}
			& $\mathcal{I}_L$ & $\mathcal{I}_R$ \\
			\hline
			bi-adjoint scalar & $\mathcal{C}_n (\omega)$ & $\mathcal{C}_n (\tilde\omega)$ \\
			Yang-Mills & $\mathcal{C}_n(\omega)$ & $\pf' \Psi_n$ \\
			Einstein gravity & $\pf' \Psi_n$ & $\pf' \tilde{\Psi}_n$ \\
			Born-Infeld & $(\pf' \A_n)^2$ & $\pf' \Psi_n$ \\
			Non-linear sigma model & $\mathcal{C}_n(\omega)$ & $(\pf' \A_n)^2$ \\
			Yang-Mills-scalar & $\mathcal{C}_n(\omega)$ & $\pf\, \textsf{X}_n \, \pf' \A_n$ \\
			Einstein-Maxwell-scalar & $\pf\, \textsf{X}_n \, \pf' \A_n$ & $\pf\, \textsf{X}_n \, \pf' \A_n$ \\
			Dirac-Born-Infeld (scalar) & $(\pf' \A_n)^2$ & $\pf\, \textsf{X}_n \, \pf' \A_n$ \\
			special Galileon & $(\pf' \A_n)^2$ & $(\pf' \A_n)^2$ \\
		\end{tabular}
	\end{center}
	\caption{\label{tab:theories}Form of the integrands for various theories}
\end{table}
Let us describe each ingredient in turn. The Parke-Taylor factor for ordering $\omega$ is given by
\be
\mathcal{C}_n (\omega) = \frac{1}{\sigma_{\omega_1 \omega_2} \cdots \sigma_{\omega_{n-1} \omega_n} \sigma_{\omega_n \omega_1}},
\en
where $T^a$ denote the generators of the group $U(N)$.

We define the $n \times n$ matrices through:
\be
[\A_n]_{ab} = \begin{dcases} 
	\frac{2 k_a \cdot k_b}{\sigma_{ab}} & a \neq b, \\
	0 & a = b.
\end{dcases}
\quad\quad
[\textsf{B}_n]_{ab} = \begin{dcases} 
	\frac{2 \epsilon_a \cdot \epsilon_b}{\sigma_{ab}} & a \neq b, \\
	0 & a = b.
\end{dcases}\nonumber
\en
\be
[\textsf{C}_n]_{ab} = \begin{dcases} 
	\frac{2 \epsilon_a \cdot k_b}{\sigma_{ab}} & a \neq b, \\
	-\sum_{c \neq a} \frac{2 \epsilon_a \cdot k_c}{\sigma_{ac}} & a = b.
\end{dcases}
\quad\quad
[\textsf{X}_n]_{ab} = \begin{dcases} 
	\frac{1}{\sigma_{ab}} & a \neq b, \\
	0 & a = b.
\end{dcases}
\en
and the $2n \times 2n$ matrix as
\be
\Psi_n =
\left[
\begin{array}{c c}
	\A_n & -\textsf{C}_n^T \\
	\textsf{C}_n & \textsf{B}_n
\end{array}
\right].
\en
The reduced Pfaffian is defined as $\pf' \Psi_n = \frac{(-)^{a+b}}{\sigma_{ab}} \pf \Psi_n^{[a,b]}$, where the notation $\Psi_n^{[a,b]}$ means the columns and rows $a,b$ of the matrix $\Psi_n$ have been removed. One can show this  definition is independent of the choice of $a,b$. Analogous notation holds for $\pf' \A_n$.

For more details about the theories defined through the CHY formula, as well as their relations please refer to \cite{CHY}.

\section{\label{sec:large z}Large-$z$ Behavior of NLSM}

In this appendix we study in turn the large-$z$ scaling of NLSM and mixed NLSM-biadjoint amplitudes under a 3-shift $(ijn)$ defined by (\ref{eq:3-shift}).

\subsection*{Non-linear Sigma Model Amplitudes}

Vertices of $U(N)$ NLSM have been studied in \cite{Kampf:2013vha}. Independently of parametrization, they are linear in the Mandelstam invariants. This can also be easily seen from the CHY formula. Knowing that the puncture locations $\sigma$ are dimensionless, the measure scales as $s^{3-n}$, the Parke-Taylor is constant, and the reduced Pfaffian factor goes as $s^{n/2-1}$. Hence the whole amplitude needs to scale as $s^1$ for any number of particles, and so must every vertex.

\begin{figure}[!ht]
	\centering
	\includegraphics[width=0.6\textwidth]{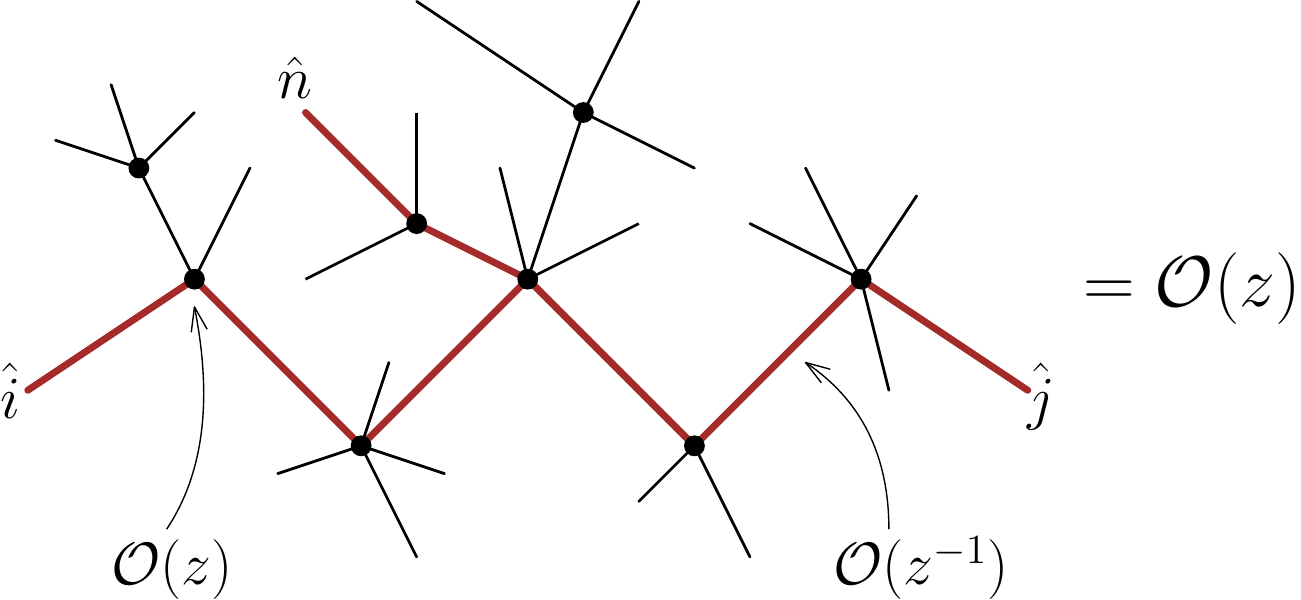}
	\caption{\label{fig:Diagram NLSM}Example diagram for a NLSM amplitude. Propagators in bold red define a subtree with deformed momenta. All vertices scale as a single power of $z$.}
\end{figure}

A 3-shift $(ijn)$ uniquely defines a subtree in the diagram, see Figure \ref{fig:Diagram NLSM}. All internal momenta in this subgraph are deformed linearly in the parameter $z$, and others stay constant. Since in the large-$z$ limit each propagator contributes as $z^{-1}$ and each vertex as $z^1$, the whole amplitude scales as $\mathcal{O}(z)$.

Strictly speaking, the above analysis provides an upper bound on this scaling. We have checked analytically up to $n=10$ that this bound is saturated whenever all of $i,j,n$ are consecutive or when two of them are adjacent. The former is the case we consider in Section \ref{sec:NLSM}.

Curiously enough, in the case when all three of the deformed momenta are non-adjacent, the amplitude seems to behave as $\mathcal{O}(1)$ in the large-$z$ limit. If this holds true, one can define yet another recursion relation scheme, this time considering the contour integral,
\be
A^{\rm NLSM}_n(\mathbb{I}_n) = -\oint_{|z|=\epsilon} dz\; \frac{\hat{A}^{\mathrm{NLSM}}_n(\mathbb{I}_n; z)}{z(1-z)},
\en
with a single factor of $(1-z)$ in the denominator. Due to the enhanced large-$z$ behavior, there are no poles at infinity.

For instance, the 6-point amplitude with 3-shift $(135)$ gets contributions from three factorization channels. Using the 4-point amplitude as a seed, we obtain:
\bes
A_6 (1^\Sigma, 2^\Sigma, 3^\Sigma, 4^\Sigma, 5^\Sigma, 6^\Sigma) &=&\frac{(s_{\hat{1}2}+s_{2\hat{3}})(s_{4\hat{5}}+s_{\hat{5}6})}{(1-z)s_{123}}\Bigg|_{z=z_{123}} + \frac{(s_{2\hat{3}}+s_{\hat{3}4})(s_{\hat{5}6}+s_{6\hat{1}})}{(1-z)s_{234}}\Bigg|_{z=z_{234}} \tr
&+&\frac{(s_{\hat{3}4}+s_{4\hat{5}})(s_{6\hat{1}}+s_{\hat{1}2})}{(1-z)s_{345}}\Bigg|_{z=z_{345}}.
\ens
This expression can be shown to give the same result as (\ref{eq:NLSM 6-pt}). We can however no longer make use of the trick (\ref{eq:trick}) in order to quickly obtain the final answer in a nice form.

\subsection*{Non-linear Sigma Model and Biadjoint Scalar Mixed Amplitudes}

To complete the analysis of the recursion relation, we need to study the behavior at infinity of an amplitude with three biadjoint scalars and $n-3$ NLSM Goldstone bosons. It is possible to do this without the knowledge of the precise form of the interactions.

First, let us remind that the formula
\be
\oint d\mu_n~\mathcal{C}_n (\mathbb{I}_n)~(\pf' \A_n)^2 = -\oint d\mu_n~\mathcal{C}_n (\mathbb{I}_n)~\frac{1}{\sigma_{ab}\sigma_{ba}}(\pf\, \A^{[ab]}_{n-2})^2
\en
can be understood (up to a sign) as either the amplitude for pure NLSM, or the one for two biadjoint scalars, $a,b$ and rest NLSM particles. It follows that all vertices with two $\phi$'s and arbitrary number of sigmas have the same form as the original NLSM vertices! Large-$z$ scaling therefore remains unmodified.

Following the previous analysis we also have that a mixed amplitude with $k$ biadjoint scalars behaves as $s^{3-k}$, independent of the total number of NLSM particles involved (as soon as it is non-zero). For the case of interest, $k=3$, this implies that the corresponding vertices in a local field theory are independent of the kinematic invariants and hence suffer no scaling in $z$.

\begin{figure}[!ht]
	\centering
	\includegraphics[width=0.6\textwidth]{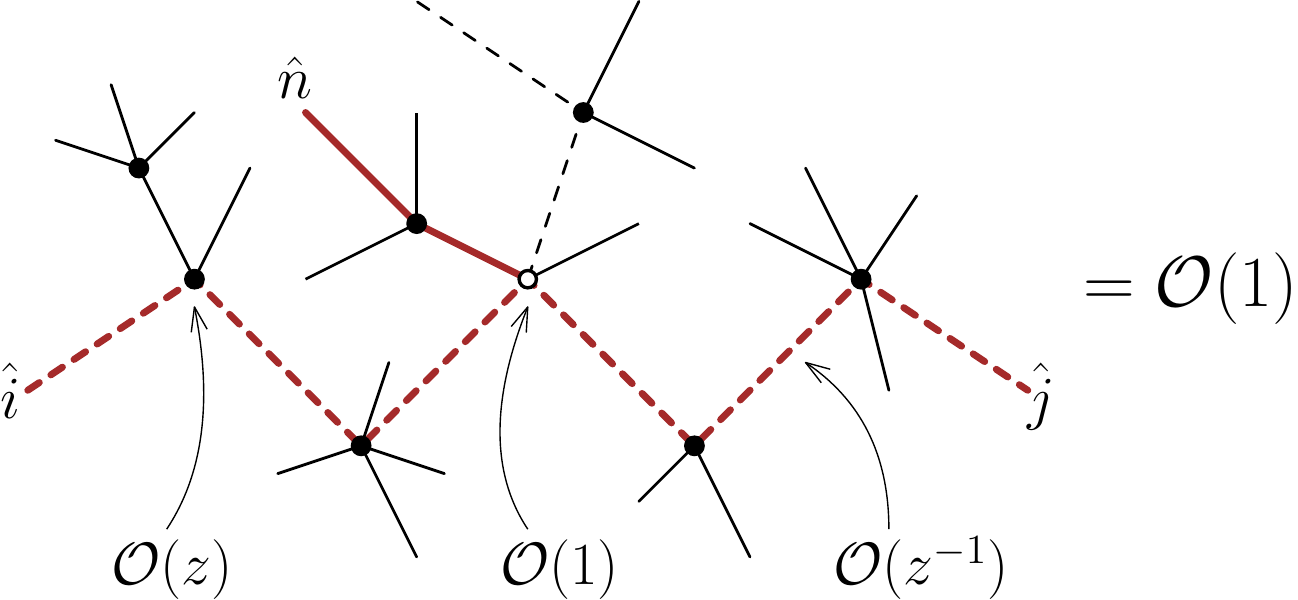}
	\caption{\label{fig:Diagram mixed}Example diagram for a NLSM $\oplus\, \phi^3$ amplitude. Solid and dashed lines define $\Sigma$ and $\phi$ propagators respectively. Propagators in bold red define a subtree with deformed momenta. Black vertices scale as $z$, while the white one remains constant.}
\end{figure}
It is now a straightforward task to obtain the scaling of the whole amplitude. We perform the 3-shift $(ijn)$ with $i,j$ being the biadjoint scalars and $n$ a single $\Sigma$. The $U(\tilde{N})$ structure carried by $\phi$'s connects them into a subtree of the diagram, see Figure \ref{fig:Diagram mixed}. The additional Goldstone boson $n$ pinches this tree in a single place, either in a vertex with two or three scalars. In either case, only one vertex along the path needs to have a constant scaling, while the remaining ones go as $z^1$, so the whole amplitude must scale as $\mathcal{O}(1)$. We numerically confirmed for $n \leq 9$ that this upper bound is in fact saturated.


\begin{thebibliography}{9}

\bibitem{Adler:1964um}
S.~L.~Adler,
``Consistency conditions on the strong interactions implied by a partially conserved axial vector current,''
Phys.\ Rev.\  {\bf 137}, B1022 (1965).
doi:10.1103/PhysRev.137.B1022

\bibitem{Susskind:1970gf}
L.~Susskind and G.~Frye,
``Algebraic aspects of pionic duality diagrams,''
Phys.\ Rev.\ D {\bf 1}, 1682 (1970).
doi:10.1103/PhysRevD.1.1682

\bibitem{Cheung:2014dqa}
C.~Cheung, K.~Kampf, J.~Novotny and J.~Trnka,
``Effective Field Theories from Soft Limits of Scattering Amplitudes,''
Phys.\ Rev.\ Lett.\  {\bf 114}, no. 22, 221602 (2015)
doi:10.1103/PhysRevLett.114.221602
[arXiv:1412.4095 [hep-th]].

\bibitem{Cheung:2015ota}
C.~Cheung, K.~Kampf, J.~Novotny, C.~H.~Shen and J.~Trnka,
``On-Shell Recursion Relations for Effective Field Theories,''
Phys.\ Rev.\ Lett.\  {\bf 116}, no. 4, 041601 (2016)
doi:10.1103/PhysRevLett.116.041601
[arXiv:1509.03309 [hep-th]].

\bibitem{ArkaniHamed:2008gz}
N.~Arkani-Hamed, F.~Cachazo and J.~Kaplan,
``What is the Simplest Quantum Field Theory?,''
JHEP {\bf 1009}, 016 (2010)
doi:10.1007/JHEP09(2010)016
[arXiv:0808.1446 [hep-th]].

\bibitem{1307.2199 CHY}
F.~Cachazo, S.~He and E.~Y.~Yuan,
``Scattering of Massless Particles in Arbitrary Dimensions,''
Phys.\ Rev.\ Lett.\  {\bf 113}, no. 17, 171601 (2014)
[arXiv:1307.2199 [hep-th]].

\bibitem{Cachazo:2013iea}
F.~Cachazo, S.~He and E.~Y.~Yuan,
``Scattering of Massless Particles: Scalars, Gluons and Gravitons,''
JHEP {\bf 1407}, 033 (2014)
[arXiv:1309.0885 [hep-th]].

\bibitem{Dolan:2013isa} 
L.~Dolan and P.~Goddard,
``Proof of the Formula of Cachazo, He and Yuan for Yang-Mills Tree Amplitudes in Arbitrary Dimension,''
JHEP {\bf 1405}, 010 (2014)
doi:10.1007/JHEP05(2014)010
[arXiv:1311.5200 [hep-th]].

\bibitem{CHY}
F.~Cachazo, S.~He and E.~Y.~Yuan,
``Scattering Equations and Matrices: From Einstein To Yang-Mills, DBI and NLSM,''
JHEP {\bf 1507}, 149 (2015)
[arXiv:1412.3479 [hep-th]].

\bibitem{Witten}
E.~Witten,
``Perturbative gauge theory as a string theory in twistor space,''
Commun.\ Math.\ Phys.\  {\bf 252}, 189 (2004)
[hep-th/0312171].

\bibitem{RSV}
R.~Roiban, M.~Spradlin and A.~Volovich,
``A Googly amplitude from the B model in twistor space,''
JHEP {\bf 0404}, 012 (2004)
[hep-th/0402016].

\bibitem{Cachazo:2012kg} 
F.~Cachazo and D.~Skinner,
``Gravity from Rational Curves in Twistor Space,''
Phys.\ Rev.\ Lett.\  {\bf 110}, no. 16, 161301 (2013)
doi:10.1103/PhysRevLett.110.161301
[arXiv:1207.0741 [hep-th]].

\bibitem{Cachazo:2012pz} 
F.~Cachazo, L.~Mason and D.~Skinner,
``Gravity in Twistor Space and its Grassmannian Formulation,''
SIGMA {\bf 10}, 051 (2014)
doi:10.3842/SIGMA.2014.051
[arXiv:1207.4712 [hep-th]].

\bibitem{Volkov:1973ix}
D.~V.~Volkov and V.~P.~Akulov,
``Is the Neutrino a Goldstone Particle?,''
Phys.\ Lett.\ B {\bf 46}, 109 (1973).
doi:10.1016/0370-2693(73)90490-5

\bibitem{He:2016vfi} 
S.~He, Z.~Liu and J.~B.~Wu,
``Scattering Equations, Twistor-string Formulas and Double-soft Limits in Four Dimensions,''
arXiv:1604.02834 [hep-th].

\bibitem{Bergshoeff:2013pia}
E.~Bergshoeff, F.~Coomans, R.~Kallosh, C.~S.~Shahbazi and A.~Van Proeyen,
``Dirac-Born-Infeld-Volkov-Akulov and Deformation of Supersymmetry,''
JHEP {\bf 1308}, 100 (2013)
doi:10.1007/JHEP08(2013)100
[arXiv:1303.5662 [hep-th]].

\bibitem{Luo:2015tat}
H.~Luo and C.~Wen,
``Recursion relations from soft theorems,''
JHEP {\bf 1603}, 088 (2016)
doi:10.1007/JHEP03(2016)088
[arXiv:1512.06801 [hep-th]].

\bibitem{GellMann:1960np}
M.~Gell-Mann and M.~Levy,
``The axial vector current in beta decay,''
Nuovo Cim.\  {\bf 16}, 705 (1960).
doi:10.1007/BF02859738

\bibitem{Kampf:2013vha}
K.~Kampf, J.~Novotny and J.~Trnka,
``Tree-level Amplitudes in the Nonlinear Sigma Model,''
JHEP {\bf 1305}, 032 (2013)
doi:10.1007/JHEP05(2013)032
[arXiv:1304.3048 [hep-th]].

\bibitem{Cachazo:2013gna}
F.~Cachazo, S.~He and E.~Y.~Yuan,
``Scattering equations and Kawai-Lewellen-Tye orthogonality,''
Phys.\ Rev.\ D {\bf 90}, no. 6, 065001 (2014)
[arXiv:1306.6575 [hep-th]].

\bibitem{Britto:2004ap} 
R.~Britto, F.~Cachazo and B.~Feng,
``New recursion relations for tree amplitudes of gluons,''
Nucl.\ Phys.\ B {\bf 715}, 499 (2005)
doi:10.1016/j.nuclphysb.2005.02.030
[hep-th/0412308].

\bibitem{Britto:2005fq} 
R.~Britto, F.~Cachazo, B.~Feng and E.~Witten,
``Direct proof of tree-level recursion relation in Yang-Mills theory,''
Phys.\ Rev.\ Lett.\  {\bf 94}, 181602 (2005)
doi:10.1103/PhysRevLett.94.181602
[hep-th/0501052].

\bibitem{Bern:2005hs} 
Z.~Bern, L.~J.~Dixon and D.~A.~Kosower,
``On-shell recurrence relations for one-loop QCD amplitudes,''
Phys.\ Rev.\ D {\bf 71}, 105013 (2005)
doi:10.1103/PhysRevD.71.105013
[hep-th/0501240].

\bibitem{Risager:2005vk} 
K.~Risager,
``A Direct proof of the CSW rules,''
JHEP {\bf 0512}, 003 (2005)
doi:10.1088/1126-6708/2005/12/003
[hep-th/0508206].

\bibitem{Hinterbichler:2011tt}
K.~Hinterbichler,
``Theoretical Aspects of Massive Gravity,''
Rev.\ Mod.\ Phys.\  {\bf 84}, 671 (2012)
doi:10.1103/RevModPhys.84.671
[arXiv:1105.3735 [hep-th]].

\bibitem{Dvali:2000hr}
G.~R.~Dvali, G.~Gabadadze and M.~Porrati,
``4-D gravity on a brane in 5-D Minkowski space,''
Phys.\ Lett.\ B {\bf 485}, 208 (2000)
doi:10.1016/S0370-2693(00)00669-9
[hep-th/0005016].

\bibitem{Kampf:2014rka}
K.~Kampf and J.~Novotny,
``Unification of Galileon Dualities,''
JHEP {\bf 1410}, 006 (2014)
doi:10.1007/JHEP10(2014)006
[arXiv:1403.6813 [hep-th]].

\bibitem{Hinterbichler:2015pqa}
K.~Hinterbichler and A.~Joyce,
``Hidden symmetry of the Galileon,''
Phys.\ Rev.\ D {\bf 92}, no. 2, 023503 (2015)
doi:10.1103/PhysRevD.92.023503
[arXiv:1501.07600 [hep-th]].

\bibitem{KLT}
H.~Kawai, D.~C.~Lewellen and S.~H.~H.~Tye,
``A Relation Between Tree Amplitudes of Closed and Open Strings,''
Nucl.\ Phys.\ B {\bf 269}, 1 (1986).

\bibitem{BjerrumBohr:2010hn}
N.~E.~J.~Bjerrum-Bohr, P.~H.~Damgaard, T.~Sondergaard and P.~Vanhove,
``The Momentum Kernel of Gauge and Gravity Theories,''
JHEP {\bf 1101}, 001 (2011)
doi:10.1007/JHEP01(2011)001
[arXiv:1010.3933 [hep-th]].

\bibitem{Du:2014eca}
Y.~J.~Du, B.~Feng, C.~H.~Fu and Y.~Wang,
``Note on Soft Graviton theorem by KLT Relation,''
JHEP {\bf 1411}, 090 (2014)
doi:10.1007/JHEP11(2014)090
[arXiv:1408.4179 [hep-th]].

\bibitem{Kallosh:1997aw}
R.~Kallosh,
``Volkov-Akulov theory and D-branes,''
Lect.\ Notes Phys.\  {\bf 509}, 49 (1998)
doi:10.1007/BFb0105228
[hep-th/9705118].

\bibitem{Chen:2014xoa} 
W.~M.~Chen, Y.~t.~Huang and C.~Wen,
``New Fermionic Soft Theorems for Supergravity Amplitudes,''
Phys.\ Rev.\ Lett.\  {\bf 115}, no. 2, 021603 (2015)
doi:10.1103/PhysRevLett.115.021603
[arXiv:1412.1809 [hep-th]].

\bibitem{Cachazo:2013iaa}
F.~Cachazo, S.~He and E.~Y.~Yuan,
``Scattering in Three Dimensions from Rational Maps,''
JHEP {\bf 1310}, 141 (2013)
[arXiv:1306.2962 [hep-th]].

\bibitem{Eulerian}
https://en.wikipedia.org/wiki/Eulerian\_number

\bibitem{Cachazo:2013zc}
F.~Cachazo,
``Resultants and Gravity Amplitudes,''
arXiv:1301.3970 [hep-th].

\bibitem{Liu:2014vva}
Z.~W.~Liu,
``Soft theorems in maximally supersymmetric theories,''
Eur.\ Phys.\ J.\ C {\bf 75}, no. 3, 105 (2015)
doi:10.1140/epjc/s10052-015-3304-1
[arXiv:1410.1616 [hep-th]].

\bibitem{Larkoski:2014bxa} 
A.~J.~Larkoski, D.~Neill and I.~W.~Stewart,
``Soft Theorems from Effective Field Theory,''
JHEP {\bf 1506}, 077 (2015)
doi:10.1007/JHEP06(2015)077
[arXiv:1412.3108 [hep-th]].

\bibitem{Tseytlin:1986ti} 
A.~A.~Tseytlin,
``Vector Field Effective Action in the Open Superstring Theory,''
Nucl.\ Phys.\ B {\bf 276}, 391 (1986)
Erratum: [Nucl.\ Phys.\ B {\bf 291}, 876 (1987)].
doi:10.1016/0550-3213(86)90303-2

\bibitem{Benincasa:2007xk} 
P.~Benincasa and F.~Cachazo,
``Consistency Conditions on the S-Matrix of Massless Particles,''
arXiv:0705.4305 [hep-th].

\bibitem{Bondi:1962px} 
H.~Bondi, M.~G.~J.~van der Burg and A.~W.~K.~Metzner,
``Gravitational waves in general relativity. 7. Waves from axisymmetric isolated systems,''
Proc.\ Roy.\ Soc.\ Lond.\ A {\bf 269}, 21 (1962).
doi:10.1098/rspa.1962.0161

\bibitem{Sachs:1962wk} 
R.~K.~Sachs,
``Gravitational waves in general relativity. 8. Waves in asymptotically flat space-times,''
Proc.\ Roy.\ Soc.\ Lond.\ A {\bf 270}, 103 (1962).
doi:10.1098/rspa.1962.0206

\bibitem{Strominger:2013jfa} 
A.~Strominger,
``On BMS Invariance of Gravitational Scattering,''
JHEP {\bf 1407}, 152 (2014)
doi:10.1007/JHEP07(2014)152
[arXiv:1312.2229 [hep-th]].

\bibitem{Amplitudes 2015}
F.~Cachazo,
``S-Matrix Theory,''
Talk at the Amplitudes 2015 Conference.
http://amp15.itp.phys.ethz.ch/talks/Cachazo.pdf

\bibitem{Barnich:2015uva} 
G.~Barnich and B.~Oblak,
``Notes on the BMS group in three dimensions: II. Coadjoint representation,''
JHEP {\bf 1503}, 033 (2015)
doi:10.1007/JHEP03(2015)033
[arXiv:1502.00010 [hep-th]].

\end{thebibliography}
\end{document}